\documentclass[aps,pre,showpcs,amsmath,amssymb,superscriptaddress,reprint]{revtex4-1}

\setcitestyle{super}

\usepackage{graphicx}
\usepackage{caption}
\usepackage{subcaption}
\usepackage{dcolumn}
\usepackage{bm}
\usepackage{hyperref}

\begin{document}
\author{Sebastian J. Davis}
\author{Michal Macha}
\author{Andrey Chernev}
\affiliation{Laboratory of Nanoscale Biology, Institute of Bioengineering, School of Engineering, EPFL, 1015 Lausanne, Switzerland}
\author{David M. Huang}
\affiliation{Department of Chemistry, School of Physical Sciences, The University of Adelaide, Adelaide, Australia}
\author{Aleksandra Radenovic}
\email{sanjin.marion@epfl.ch, aleksandra.radenovic@epfl.ch}
\author{Sanjin Marion}
\email{sanjin.marion@epfl.ch, aleksandra.radenovic@epfl.ch}
\affiliation{Laboratory of Nanoscale Biology, Institute of Bioengineering, School of Engineering, EPFL, 1015 Lausanne, Switzerland}

\title[]{Pressure Induced Enlargement and Ionic Current Rectification in Symmetric Nanopores}


\begin{abstract}
Nanopores in solid state membranes are a tool able to probe nanofluidic phenomena, or can act as a single molecular sensor. They also have diverse applications in filtration, desalination, or osmotic power generation. Many of these applications involve chemical, or hydrostatic pressure differences which act on both the supporting membrane, and the ion transport through the pore. By using pressure differences between the sides of the membrane, and an alternating current approach to probe ion transport, we investigate two distinct physical phenomena: the elastic deformation of the membrane through the measurement of strain at the nanopore, and the growth of ionic current rectification with pressure due to pore entrance effects.
\end{abstract}

\maketitle

\section{Introduction}

Nanopores are a single molecule tool with diverse applications in bio-sensing,\citep{Plesa2013,Merchant2010} osmotic power generation\citep{Macha2019} and water desalination.\citep{Noy2020} A nanoscale pore separates two reservoirs filled with electrolyte. Monitoring ion transport through the pore yields information about a passing analyte such as DNA, or on non-linear phenomena such as ionic current rectification (ICR)\citep{Siwy2006} and other nanofluidic effects.\citep{Gravelle2019,Poggioli2019,Bocquet2010} Solid state nanopores are readily made in silicon nitride suspended membranes since they are compatible with standard lithography techniques. Pores in these suspended membranes can be used as such, as in this study, or can further support a membrane made of quasi-2D materials such as molybdenum disulphide, hexagonal boron nitride, or graphene in which a small pore can be further drilled.\citep{Graf2019}


The combination of hydrostatic pressure gradients with nanopores has so far been mostly used to study analyte translocations,\citep{Zhang2013, Lu2013,Li2017,Hoogerheide2014} the surface charge of the pore,\citep{Firnkes2010} or as a tool to control wetting.\citep{Marion2019}. It has been shown that pressure can strongly influence the ion transport properties of a nanopore or nanochannel depending on the system's resistance to hydraulic fluid flow, and modulate ion transport.\citep{Mouterde2019,Marcotte2020} On the other hand, ionic current rectification,\citep{Poggioli2019} which is linked to ion selectivity, has been found to be reduced in conical pores under the influence of pressure induced fluid flow.\citep{Lan2011,Jubin2018}



The application of pressure on thin supported membranes is a well established technique for studying the elastic properties of thin films. Blistering of thin membranes such as silicon nitride,\citep{vlassak1992} or blistering and delamination of 2D materials\citep{Koenig2011,Boddeti2013,Bunch2008} has been extensively studied in dry conditions. Studies in liquid and with nanopores have so far been restricted to nanopores drilled in elastomeric membranes for studying analyte translocations.\citep{Willmott2008,Roberts2010} No experiments have been performed so far with nanopores in elastic solid-state membranes, although such membranes are usually used in conditions of osmotic or hydraulic pressure gradients which could influence pore properties like ion selectivity and water permeability.\cite{Humplik2011,Macha2019,Noy2020} Theoretical work on sub-nm pores in 2D materials indicates the presence of strong mechanosensitivity to lateral stresses.\citep{Weifeng2016,Fang2019,Sahu2019,Fang2019b,Smolyanitsky2020}
To realize a truly mechanosensitive solid-state sensor, one which would mimic mechanosensitive biological channels,\cite{Cox2019} one needs to first understand the elastic behaviour of nanopores in solid state membranes.

This study aims to quantify the role of hydraulic pressure in modulating ion transport in thin symmetrical nanopores using a phase sensitive amplifier enhancing the sensitivity. We decouple two independent physical phenomena. First, that the pressure induced deformation of the supporting membrane causes an enlargement in the nanopore size. This allows direct measurement of the local membrane stress in a liquid environment as a precursor for stressing 2D material nanopores and probing mechanosensitivity.\cite{Fang2019} Secondly, we demonstrate that pressure induced-fluid flow produces ionic current rectification despite the lack of the usually required geometrical asymmetry in the pore.\cite{Siwy2006,Poggioli2019} This is opposite to the so far reported role of pressure in reducing ICR in asymmetrical nanopores\cite{Lan2011,Jubin2018}.

\section{Pressure application experimental setup}

\begin{figure*}[t!]
\centering
\includegraphics[width = \textwidth]{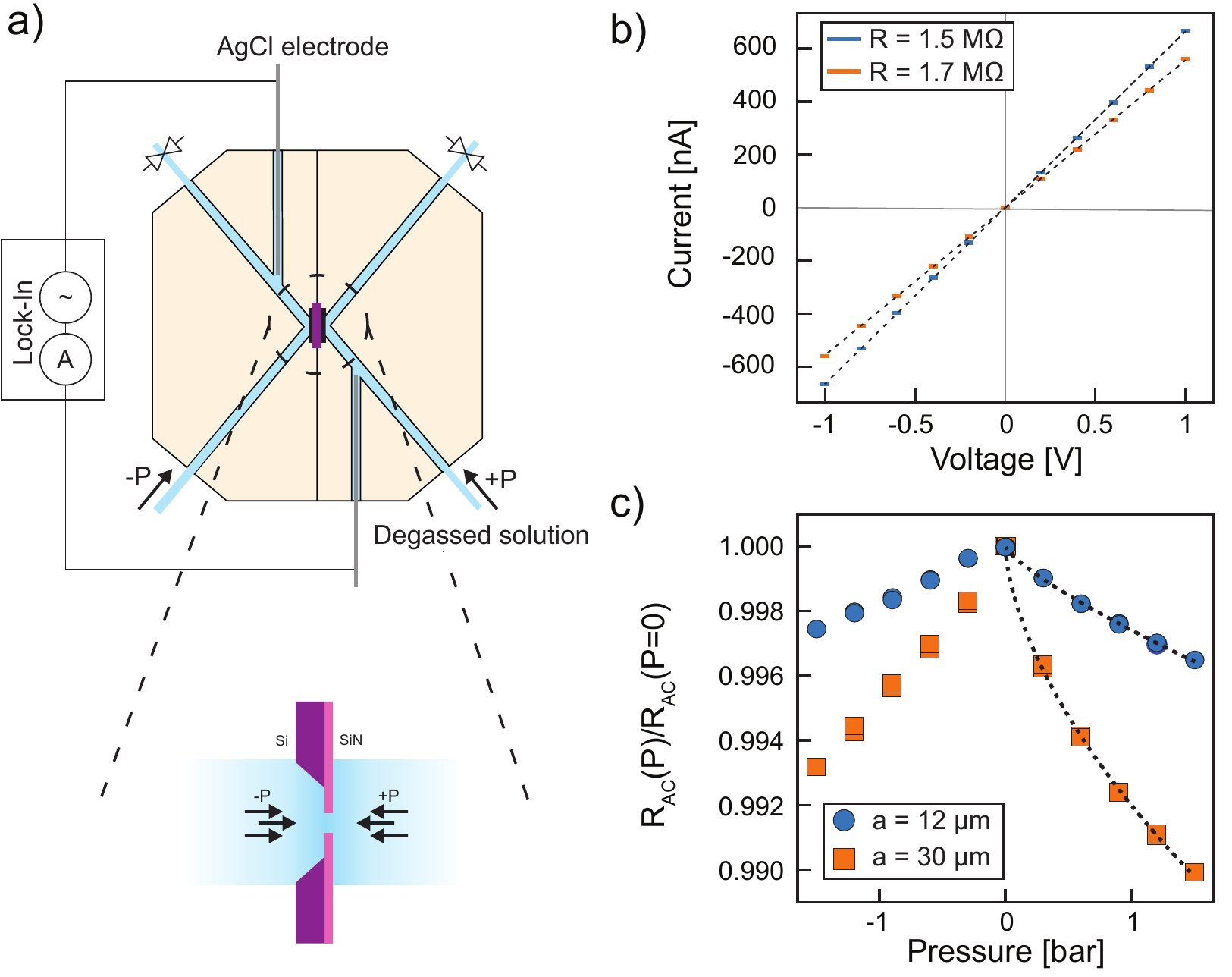}
\caption{\small \textbf{Application of pressure to solid state nanopores. a)} Schematic in side view of the sealed pressure chamber channels and electrical measurement. Zoom below shows the chip area and the convention of pressure sign chosen for this study. \textbf{b)} DC IV curves for two representative samples having two different membrane sizes but similar pore sizes of $d_0 \approx 80$ nm. Lines are linear fits giving resistance values of 1.5 M$\Omega$ and 1.7 M$\Omega$. \textbf{c)} Mean values of resistance $R_{AC}$ as a function of pressure normalized by the value of the resistance under no pressure $R_{AC}(P=0)$. The curves represent the same samples as in panel (b) with two distinct square membrane sizes of side lengths $a=12\ \mu$m and $30\ \mu$m. Lines on the positive pressure side correspond to a fit to eq.\ \ref{eq:strain} without residual stress giving $a = 12.8 \pm 1.0$ $\mu$m, and $a = 30.6 \pm 6.2$ $\mu$m respectively. The corresponding total deflection of the membrane at $P=1500$ mbar is $w_0 = 3.2 \pm 0.8$ $\mu$m, and $2.7 \pm 0.6$ $\mu$m respectively.}
\label{fig:setup}
\end{figure*}

To study how hydrostatic pressure and potential differences influence the ion transport through a solid state nanopore we use a sealed, pressure-tight chamber (See Materials and Methods and figure\ \ref{fig:setup}a) as described previously.\cite{Marion2019} After a sample consisting of a membrane with an $80$ nm diameter nanopore is mounted into the chamber, the system is wetted with a degassed $1$ M KCL buffered solution under $7$ bar compression pressure. Pressure $P$ is applied on the sample membrane using a microfluidics pressure controller. We define positive pressure as being applied from the front-side of the membrane (flat side), and negative pressure is defined as being applied from the back-side (etch-side) (as seen on figure \ref{fig:setup}a). A potential difference $V$ between the two sides of the membrane is applied and read with Ag/AgCl electrodes. Measurements of current $I$ versus applied potential $V$ are shown on figure \ref{fig:setup}b. Only samples showing stable conductance and current noise levels over the span of the measurement were considered for further analysis (See Supporting information Sec.\ S2).

The current response of the nanopore to an external potential difference $V$, and a pressure difference $P$ between the two sides of the membrane is of the form $I = G(V, P)V + H_s P$, where $H_s$ is the streaming conductance, and $G(V,P)$ the electrical conductance. Taking into account that the non-linearity in conductance is almost negligible (figure \ref{fig:setup}b), we perform a Taylor expansion of the conductance $G(V,P) \approx G_1 (P) + G_2 (P) V $, with $G_1$ and $G_2$ corresponding to the linear and first nonlinear contribution.\cite{Marion2019} The conductance term $G_1$ has contributions from the pore interior, and the access region resistance and obeys $G_1 = \sigma\left[4L/\pi d^2 + 1/d \right]^{-1}$, where $d$ is the diameter of the nanopore, $L$ the thickness of the membrane, and $\sigma$ the bulk conductance of the solution.\citep{Hall1975,Kowalczyk2011} One measure of the nonlinearity in ion transport is the ionic current rectification (ICR) ratio\citep{Siwy2006,Poggioli2019} which we define as:
\begin{equation}
\begin{split}
r (V, P) &=\frac{| I(+V, P)-I(V=0, P)|}{| I(-V, P)-I(V=0, P)|}  \\
&\approx \frac{G_1 (P) + G_2 (P) |V|}{G_1 (P) - G_2 (P) |V|}
\end{split}
\label{eq:ICR}
\end{equation}
to exclude any streaming contribution. 

In order to deconvolute the linear and nonlinear ion transport contributions of the nanopore, and eliminate any streaming current contribution, we perform all measurements using a quasi-static AC measurement. All AC measurements are performed using a sinusoidal voltage at a frequency of $f = 1$ Hz, where the resistance matches the DC measured value and no signal leakage through parasitic chip capacitance is present.\cite{Marion2019} We use a phase sensitive amplifier, which can independently measure both $G_1$ and $G_2$ by averaging out any components of the measured current which are not at the base measurement frequency $f$ or one of its multiples. Thus the current measured with the AC voltage does not have the streaming contribution included, and we obtain the total current which has two independently measured components $I_1= G_1(P) V_\text{AC}$ and $I_2= G_2 (P) V_\text{AC}^2$, which are used to calculate the ICR ratio $r$ defined in equation \ref{eq:ICR}. We therefore perform AC measurements, with high precision, to extract the linear pore resistance $R_\text{AC}=G_1^{-1}$, and the ionic current rectification $r$ at different pressures $P$ (See Supporting information S2 for details).

\section{Strain induced pore enlargement}

\begin{figure*}[t!]
\centering
\includegraphics[width = \textwidth]{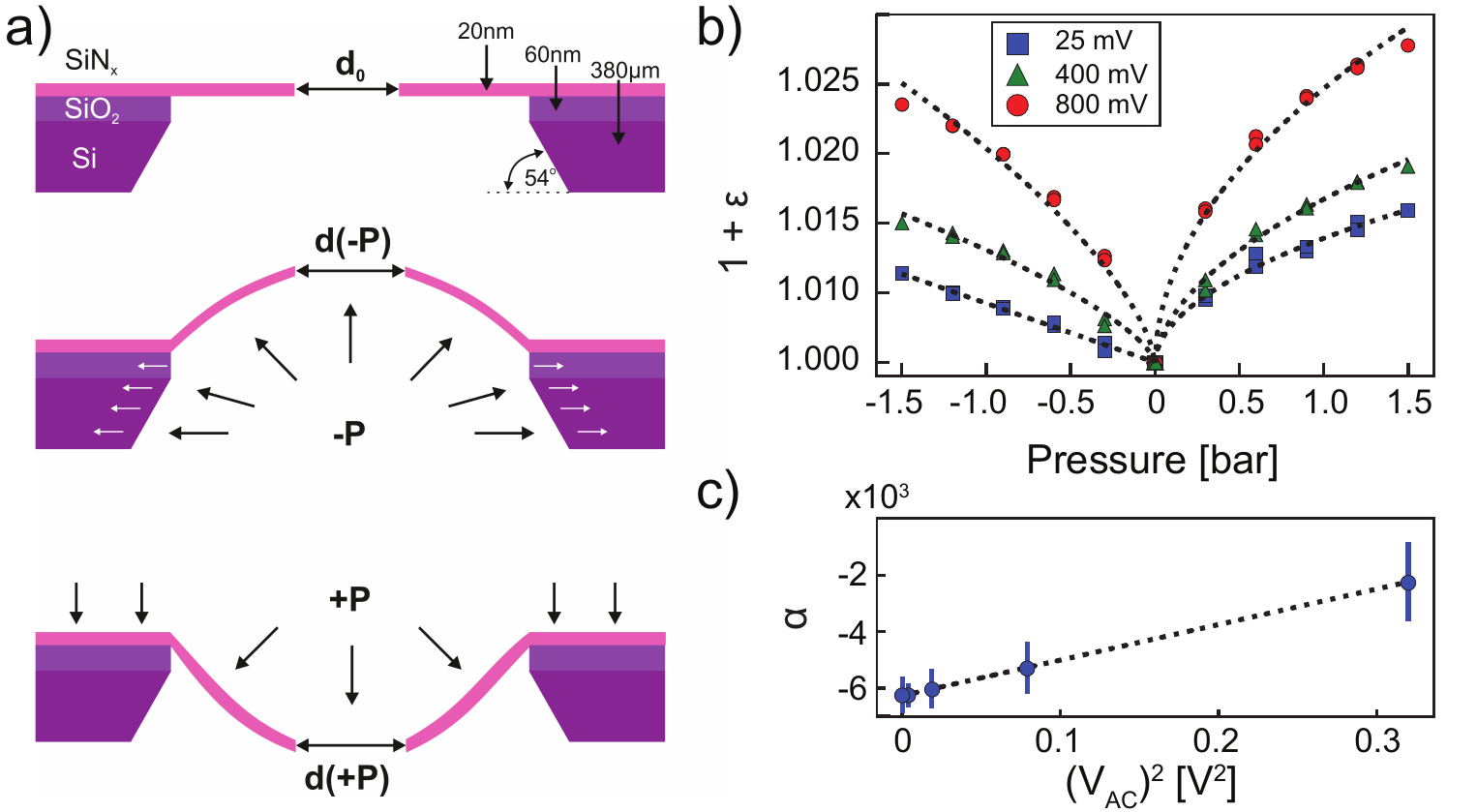}
\caption{\small \textbf{Elastic response of membranes under pressure: strain induced enlargement. 
a)} Schematic of the chip and membrane with nanopore. The initial configuration under zero applied pressure is shown as well as two schematics showing the deformation of the membrane under both positive, and negative pressure. This deformation due to strain enlarges the pore, $d(-P)$ and $d(+P)$. The negative pressure is also shown to act on the etched walls of the back-side of the chip which is responsible for the asymmetry in pressure response. \textbf{b)} Normalised strain value as a function of pressure for the same membrane at a low bias voltage of $V_\text{AC} = 25$ mV (blue squares), and two larger bias voltages of $V_\text{AC} = 400$ mV (green triangles), and $V_0 = 800$ mV (red circles). All AC voltages are given in root mean square values of the amplitude. Dashed black lines correspond to the fit of the stress to Eq.\ \eqref{eq:strain}. For positive pressures the fit is done with $\sigma_0 = 0$ fixed, and for the negative side with the full stress eq.\ \eqref{eq:strain}. \textbf{c)} Voltage dependence of the residual stress factor $\alpha$ with the dashed line corresponding to a quadratic fit to applied voltage $V_\text{AC}$. Error bars represent the standard deviation as obtained from the fit.}
\label{fig:elastic}
\end{figure*}

When pressure is applied to one side of the suspended silicon nitride membrane it blisters due to the resulting strain. Its deformation can be modelled as a thin sheet under large elastic deformations due to a uniform load in the direction perpendicular to the plane of the membrane.\citep{Timoshenko1959} The system under study has a $d_0 = 80$ nm diameter pore in the center of the square membrane of side length $a=10-30$ $\mu$m. This hole can be treated as a small perturbation which will not significantly influence the stress distribution in the membrane. As we measure the resistance of the pore $R_\text{AC}$, and this resistance is related to its diameter $d$, any change of resistance with pressure can be related with a modification of the pore diameter. The resistance decreases independently of the direction of applied pressure (figure \ref{fig:setup}c), and does not depend on the nanopore surface charge (Supporting figure S6). The measured change of pore resistance $R_\text{AC}$ with pressure is attributed to the local strain at the nanopore due to stress in the membrane. As the stress is radially symmetric at the center of the membrane, and the elastic model for the membrane involves only linear elastic deformations in the plane of the membrane, the change in size of the nanopore is trivially shown\citep{Dye2020} to be $d(P) = d_0 \left(1 + \epsilon (P) \right)$ where $\epsilon (P)=(1-\nu^2)\sigma_r (P)/E$ is the pressure dependent strain, $\sigma_r$ is the radial stress in the membrane, and $d_0$ is the pore diameter under no applied stress. Thus by precisely measuring the change in the nanopore resistance, we obtain the value of the local strain/stress at the membrane.

The elastic response of silicon nitride membranes is well studied\citep{vlassak1992,Bunch2008} which allows us to validate our model of pore enlargement. The elastic response will depend on the applied pressure $P$ as well as the geometric and elastic parameters of the membrane: $a$ the size of the square membrane, $L$ the thickness of the membrane, $E$ the Young's modulus, and $\nu$ the Poisson ratio. In addition, even under no external pressure load, the membrane will have some degree of residual stress $\sigma_0$ acting to either stretch or compress the membrane in the lateral direction. In this regime, neglecting bending of the membrane, and assuming that the stress is constant all over the membrane, the stress can be described by:\citep{vlassak1992}
\begin{equation}
\label{eq:strain}
\sigma_r^3 - \sigma_0 \sigma_r^2 - \frac{EP^2a^2}{6L^2(1-\nu)^2}= 0.
\end{equation}
By inserting the pressure dependent diameter $d(P)$ into the conductivity of the nanopore $G_1$ we are able to reproduce the dependence of the strain at the pore $\epsilon$ at different pressures. Figure \ref{fig:elastic}a shows a fit of the strain $\epsilon$ measured due to nanopore enlargement at different values of the pressure difference $P$ and applied sinusoidal voltage amplitude. The elastic parameters are taken to be $\nu = 0.23$, $L = 20$ nm, and $E = 200$ GPa, which is the average Young modulus dependent on the specifics of the fabrication procedure.\citep{Buchaillot1997} The positive pressure behaviour is fitted at a driving potential of $25$ mV to a simplified $\sigma_0 = 0$ case, while the negative pressure is fitted with the residual strain included in the fit. We find excellent agreement with the model for low electrical driving potentials, and can also correctly obtain the membrane sizes for different samples (figure \ref{fig:setup}c).

While the prediction of the membrane size shows that the simplified $\sigma_0 = 0$ case is in good agreement with the behaviour it is not sufficient to completely explain the asymmetry at low voltage in the negative pressure (as seen in figure \ref{fig:elastic}b). A fit assuming a constant $\sigma_0$ in the negative pressure direction gives values of up to 1 GPa, much higher than usually reported values of below $500$ MPa for different growth conditions,\citep{Temple-Boyer1998,Noskov1988} and not supported by the low level of deformation of the membranes we measured by atomic force microscopy (Supporting figure S4). In addition, residual stress of the membrane would affect both the positive and negative pressure behaviour and as such does not explain the observed asymmetry with pressure. We propose that the cause of this effect is due to the back side etched cavity present on the chips (figure \ref{fig:elastic}a). Application of pressure to the back-side of the chip induces forces on the etched silicon walls inside the cavity which tends to stretch the suspended membrane and modify the residual strain. Assuming a pressure dependent residual stress for the negative side of the form $\sigma_0 = \alpha P$ we find a value of $\alpha \approx -6000$ at the lowest applied sinusoidal potential for negative pressures (figure \ref{fig:elastic}c). This value can be rationalised from geometrical considerations. The applied pressure will induce a force $F_{in} \propto L_\text{Si}P \sin (54,74^{\circ} )$, where $L_\text{Si} = 380$ $\mu$m is the thickness of the silicon substrate, with the angle $54,74^{\circ}$ defined by crystallographic planes. This estimate gives a comparable induced residual stress factor of $\alpha \approx L_\text{Si}\sin(54^{\circ} )/L \approx -10000$ while neglecting any fine effects dependent on the manufacturing process.

Although including a pressure dependent residual stress on the negative pressure side explains most of the measured behaviour, figure \ref{fig:elastic}c shows that the induced residual stress factor $\alpha$ depends quadratically on the applied voltage. We propose that this effect is due to electrostriction of the underlying chip material which is known to occur for all dielectrics at high electric field regardless of crystal symmetry.\citep{Sterkenburg1992,Blaffart2013} Considering the thickness of the materials in question, the electric field at $800$ mV RMS is on the order of $2$ kV/m over the silicon substrate and on the order of $40$ MV/m over the $20$ nm thick silicon nitride membrane, sufficient to produce several percent of strain due to electrostriction. This stress counterbalances the pressure induced residual stress discussed above returning a symmetric pressure profile at high voltage. At large voltages the measured data deviates from the model and we assume that the stresses in these cases are no longer within the range of validity of eq.\ \ref{eq:strain}. 

\section{Pressure induced ionic current rectification}

\begin{figure*}
\centering
\includegraphics[width = \textwidth]{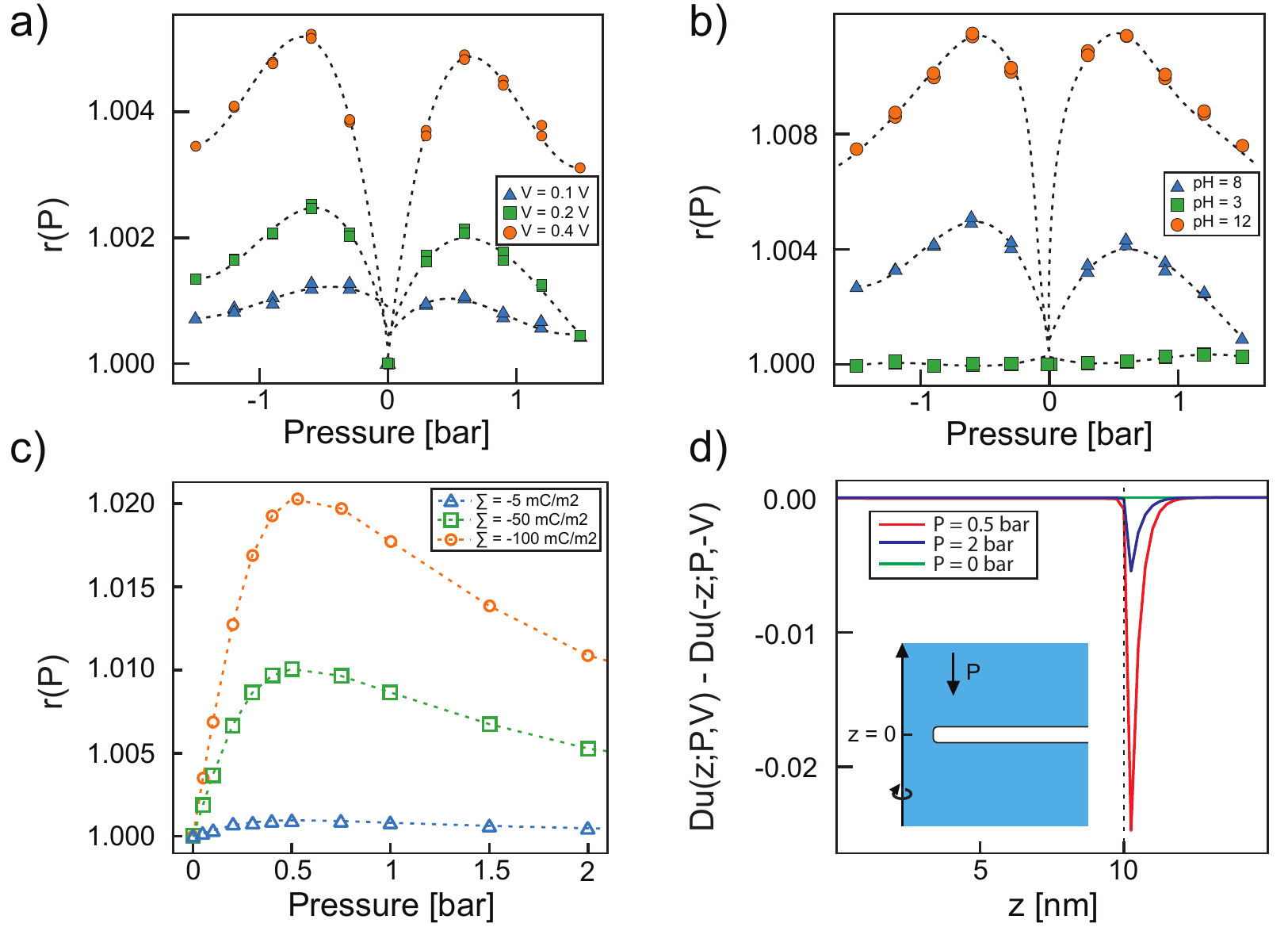}
\caption{\small \textbf{Pressure-induced rectification in symmetric solid state nanopores. a)} Rectification $r$ as a function of pressure $P$ for different driving voltages $V_\text{AC}$. The magnitude of the pressure dependent effect is seen to be approximately linear in voltage. Rectification values are corrected by pressure sweeping for baseline drift and offsets as described in the Materials and Methods. \textbf{b)} Measured rectification for three different pH values, corresponding to different surface charge densities. Corresponding streaming current measurements are provided on Supporting figure S5. \textbf{c)} Rectification extracted from a COMSOL model of a solid state nanopore under pressure. The model shows only positive pressure as it is symmetric by design. Surface charges of $-5$ mC$/$m$^2$, $-50$ mC$/$m$^2$, and $-100$ mC$/$m$^2$ are chosen to simulate the pH $3$, $8$, and $12$ case respectively. \textbf{d)} Spatial asymmetry in the Dukhin number for positive and negative bias Du$(z;P,V)-$Du$(-z;P,-V)$ along the pore axis. Three representative pressures are shown, corresponding to $P=0$, approximate maximum in ICR at $P=0.5$ bar, and in the region of ICR reduction with pressure at $P=2$ bar. Inset shows the radially symmetric FEM simulation geometry, pressure direction, and $z$ axis.}
\label{fig:rectification}
\end{figure*}

After studying pore enlargement under pressure, we investigate how pressure modifies the nonlinear conductivity of the nanopore quantified by ICR. Figure \ref{fig:rectification} shows how the ICR ratio increases with applied pressure up to a maximum value at $P \approx 500$ mbar, and then starts to reduce with a higher pressure magnitude. The decrease in ICR with an increase in pressure is well known,\citep{Lan2011,Jubin2018,Lan2016} but to our knowledge an increase in ICR with pressure has not yet been reported. The magnitude of ICR is known to be strongly dependent on the surface charge,\citep{Siwy2006,Poggioli2019} so we change its value by varying the pH of the solution. The point of zero charge for silicon nitride membranes is known to be $\approx$ pH $4$.\citep{Firnkes2010} Figure \ref{fig:rectification}b shows how a pH larger than 8 increases the ICR magnitude due to a slight increase in surface charge while not changing the pressure dependence. Conversely going near the point of zero charge at pH $3$ completely removes any pressure dependence of ICR. Here the magnitude of ICR is small as we use a high salt concentration ($1$ M KCl), but is expected to grow at lower concentrations due to a larger contribution from the surface double layer.\citep{Lee2012}

To explain the origin of the pressure induced ionic current rectification, we perform finite element method (FEM) modelling in COMSOL multiphysics. Coupled Poisson-Nernst-Planck-Stokes equations are solved with different static pressures between the two electrolyte reservoirs while varying the surface charge $\Sigma$, DC voltage bias, and pressure (See Supporting information Sec. S5). Considering the complete decoupling of the strain effect no change in shape of the pore due to the elastic deformations is considered. Figure \ref{fig:rectification}c shows the FEM values of rectification computed based on eq.\ \ref{eq:ICR} as a function of pressure for three surface charge values chosen to simulate the effect of experimental pH changes. Only positive pressure gradients are shown since the measurement is by definition symmetric in pressure. The FEM model completely captures the behaviour seen in the experimental data on figure \ref{fig:rectification}b, with an increase in $r$ at low pressures before a turnover and decrease at higher pressures. The measured decrease in magnitude of the effect as the surface charge is reduced is also captured.

The rectification behaviour can be rationalised in terms of perturbations to the ion distributions in and around the nanopore caused by pressure-induced advection. ICR in a nanapore has been shown previously to be controlled by the spatial variation in the axial direction $z$ of the local Dukhin number $\mathrm{Du}(z)$, with stronger asymmetry of $\mathrm{Du}(z)$ between the pore ends yielding stronger rectification.\cite{Poggioli2019} The Dukhin number measures the relative magnitude of surface to bulk ionic conduction. For a 1:1 electrolyte, and in the absence of Debye layer overlap in the pore, $\mathrm{Du}(z) = -\frac{\langle c_+(z) - c_-(z)\rangle}{2(c_+(z,r=0)+c_-(z,r=0))}$, where $c_\pm$ are the positive and negative ion concentrations, $\langle \cdots \rangle$ denotes an average over the pore cross-section, and $r$ is the radial coordinate.\cite{Poggioli2019} 
Pressure-driven flow induces spatial asymmetry in $\mathrm{Du}(z)$ since conservation of ion current as the bulk solution is transported into the charged nanopore perturbs both the local ionic charge density $n_c=e(c_+ - c_-)$, and local total ion concentration $c_\text{tot}=c_+ + c_-$ (Supplemental figures S8 and S9 respectively), particularly when coupled with the applied electric field. At sufficiently high pressures, however, convection completely replaces the fluid inside the nanopore by the bulk solution, reducing the spatial variation of $\mathrm{Du}(z)$ and diminishing ICR, as observed in both experiments and FEM simulations. The spatial asymmetry of $\mathrm{Du}(z)$ at positive versus negative bias $\pm V$ for different pressures $P$ from the FEM simulations is quantified by $\mathrm{Du}(z;P,V) - \mathrm{Du}(-z;P,-V)$ in figure \ref{fig:rectification}d, which confirms that the asymmetry is greatest at an intermediate pressure corresponding to the strongest rectification.

The pressure-induced asymmetry in $\mathrm{Du}(z)$ is localized to the pore ends in the FEM simulations (figure \ref{fig:rectification}d). Thus, rectification is expected to be controlled by a dimensionless P\'eclet number $\mathrm{Pe} = \frac{ud}{D}$ quantifying the relative importance of ion advection compared with diffusion in which the characteristic length scale is the pore diameter $d$. Here $u$ is the average pressure-driven fluid velocity and $D$ the diffusivity of the ions (which is approximately the same for K$^+$ and Cl$^-$). Rectification is expected to be pronounced for $\mathrm{Pe} > 1$ and to diminish as $\mathrm{Pe} \rightarrow \infty$. Consistent with this picture, the maximum rectification factor in the experiments and FEM simulations (at $P \approx$~500~mbar) occurs at $\mathrm{Pe} \approx 7$, if we take $u \approx \frac{d^2 P}{2\eta\left(16L+3\pi d\right)}$,\cite{Weissberg1962} the average fluid velocity magnitude across a nanopore of length $L$ and diameter $d$ for fluid viscosity $\eta$ due to an applied pressure $P$ and use the experimental/simulation parameter values.

\section{Conclusions}

By coupling a nanopore inside a thin elastic supported silicon nitride membrane immersed in liquid, we demonstrated how one can use AC measurements of ion transport coupled with hydrostatic pressure to precisely measure two separate physical phenomena. By monitoring the size of a nanopore while the membrane is undergoing pressure induced blistering, we demonstrate that one can precisely measure local strain in the membrane. As these membranes are typically used as a support for 2D material nanopore measurements, this can be the first step in measuring mechanosensitivity in 2D materials\citep{Weifeng2016,Fang2019,Sahu2019,Fang2019b,Smolyanitsky2020} as it allows calibration and control of applied stresses. Stress in the 2D membrane under deformation is expected to cause restructuring of the bonds in the nanopore edges, opening up a pathway for ion transport, in direct analogy to biological ion channels.\citep{Cox2019} This could provide a stress-sensitive alternative to the newly reported pressure sensitive ion transpor behaviour in single digit carbon nanotubes.\cite{Marcotte2020} In addition to strain induced enlargement of nanopores, we have shown how thin nanopores can induce nonlinear transport phenomena such as ionic current rectification. This is in contrast to the so far reported effect of the reduction of ICR with pressure.\citep{Lan2011,Jubin2018} Similar to systems which have liquid flow slippage, like long carbon nanotubes\cite{Marcotte2020}, or angstrom slits\cite{Mouterde2019}, membranes in almost-2D membranes have low hydraulic resistance which, along with access effects, produces novel nonlinear nanofluidic phenomena.

\section*{Supporting information}

The supporting information contains the Materials and methods section, details about the strain and ionic current rectification measurements and FEM model details with additional plots.

\section*{Author contributions}

S.J.D. performed the experiments, analysed the data and performed FEM simulations. S.M. designed and built the experimental set-up and built the FEM model. M.M. designed the microfluidic chamber and performed AFM imaging. A.C. fabricated devices. A.R. and S.M. supervised the research. D.M.H. provided an explanation for the ionic current rectification. S.J.D. and S.M. wrote the manuscript with all authors providing important suggestions for the experiments, discussing the results, and contributing to the manuscript.

\section*{acknowledgement}
The authors thank Marko Popovic and Alex Smolyanitsky for useful discussions on the membrane elasticity. This work was financially supported by the Swiss National Science Foundation (SNSF) Consolidator grant (BIONIC BSCGI0\_157802) and from the European Union's Horizon 2020 research and innovation programme under the Marie Skłodowska-Curie grant agreement No 754462.

\bibliographystyle{apsrev4-1}

{ \small
\bibliography{elasticity}

}

\end{document}